\documentclass[reqno,centertags,11pt]{amsart}
\usepackage{amsmath,amsthm,amscd,amssymb}
\usepackage{latexsym}

\newcommand{\bbR}{{\mathbb{R}}}

\newcommand{\bbZ}{{\mathbb{Z}}}

\newcommand{\calH}{{\mathcal H}}

\newcommand{\dott}{\,\cdot\,}

\newcommand{\lb}{\label}
\newcommand{\f}{\frac}

\newcommand{\ess}{\text{\rm{ess}}}

\newcommand{\bi}{\bibitem}

\newcommand{\beq}{\begin{equation}}
\newcommand{\eeq}{\end{equation}}
\newcommand{\ba}{\begin{align}}
\newcommand{\ea}{\end{align}}
\newcommand{\veps}{\varepsilon}
\newcommand{\Tra}{\operatorname{Tr}}
\newcommand{\Ran}{\operatorname{Ran}}

\DeclareMathOperator{\Ima}{Im}

\allowdisplaybreaks
\numberwithin{equation}{section}

\newtheorem{theorem}{Theorem}[section]
\newtheorem*{t1}{Theorem 1}
\newtheorem*{t2}{Theorem 2}
\newtheorem*{t3}{Theorem 3}
\newtheorem*{t4}{Theorem 4}
\newtheorem{proposition}[theorem]{Proposition}
\newtheorem{lemma}[theorem]{Lemma}

\theoremstyle{definition}
\newtheorem{example}[theorem]{Example}

\theoremstyle{remark}

\newcommand{\abs}[1]{\lvert#1\rvert}

\begin{document}
\title[Lieb-Thirring Inequalities for Jacobi Matrices]
{Lieb-Thirring Inequalities \\ for Jacobi Matrices}
\author[D.~Hundertmark and B.~Simon]{Dirk Hundertmark$^1$ and Barry Simon$^{1,2}$}

\thanks{$^1$ Department of Mathematics 253--37, California Institute of Technology,    
Pasadena, CA 91125, U.S.A.; E-mail: dirkh@caltech.edu, bsimon@caltech.edu} 
\thanks{$^2$ Supported in part by NSF grant DMS-9707661. }
   
\date{November 30, 2001}

\begin{abstract} 
For a Jacobi matrix $J$ on $\ell^2 (\bbZ_+)$ with $Ju(n)=a_{n-1} u(n-1) + b_n u(n) + 
a_n u(n+1)$, we prove that
\[
\sum_{\abs{E}>2} (E^2 -4)^{1/2} \leq \sum_n \abs{b_n} + 4\sum_n \abs{a_n -1}.
\]
We also prove bounds on higher moments and some related results in higher dimension.
\end{abstract}

\maketitle

\section{Introduction}\label{S1}

Let $J$ be a Jacobi matrix, that is, a tridiagonal matrix 
\[
J= \begin{pmatrix} 
b_1 & a_1 & 0 & 0 & \dots \\
a_1 & b_2 & a_2 & 0 & \dots \\
0 & a_2 & b_3 & a_3 & \dots \\
0 & 0   & a_3 & b_4 & \dots \\
\vdots & \vdots & \vdots & \vdots & \ddots 
\end{pmatrix}
\]
viewed as an operator on $\ell^2 (\bbZ_+)$ via 
\begin{equation} \lb{1.1}
(Ju)(n) = a_{n-1} u(n-1) + b_n u(n) + a_n u(n).
\end{equation}
Here $a_n >0$ and $b_n\in\bbR$. We will sometimes denote the variables in $J$ explicitly 
by writing $J(\{a_n\}_{n\geq 1}, \{b_n\}_{n\geq 1})$. We are interested in perturbations  
of the special case $a_n \equiv 1$, $b_n=0$, called $J_0$, the free Jacobi matrix and, in 
particular, the case where $J-J_0$ is compact, viz. $a_n\to 1$, $b_n\to 0$ as $n\to\infty$. 
Then $\sigma_{\ess} (J)=\sigma_{\ess}(J_0)=[-2,2]$ and $J$ has simple eigenvalues 
$\{E_n^\pm\}_{n=0}^{N_\pm}$ with ($N_+$ or $N_-$ or both may be infinite) 
\begin{equation} \label{1.2}
E_1^+ > E_2^+ > \cdots > 2 > -2 >\cdots > E_2^- >E_1^-.
\end{equation}

One of our main goals in this paper is to prove the bound

\begin{t1} 
\begin{equation} \label{1.3}
\sum_{n=1,\dots, N_\pm} [(E_n^+)^2 -4]^{1/2} + [(E_n^-)^2 -4]^{1/2} \leq \sum_n 
\abs{b_n} + 4\sum_n \abs{a_n -1}
\end{equation}
\end{t1}

As we will see, the constants $1$ in front of the $b$ sum and $4$ in the $a_n-1$ sum are 
both optimal. \eqref{1.3} is optimal in another regime, namely, large coupling for $b$. 
Specifically, let $J_\lambda$ be defined with $a_n =a_n^{(0)}$ and $b_n = \lambda b_n^{(0)}$. 
Let $\tilde b_n^\pm$ be a reordering of the $b_n$'s with $\pm \tilde b_n^\pm >0$ so 
$\tilde b_1^+ \geq \tilde b_2^+ \geq\cdots\geq 0$ and $\tilde b_1^- \leq \tilde  b_2^- 
\leq\cdots\leq 0$. Then it is not hard to see that 
\begin{equation} \lb{1.3a} 
\lim_{\lambda\to\infty} \, \lambda^{-1} E_n^\pm (J_\lambda) = \tilde b_n^\pm
\end{equation} 
which shows that the ratio of the two sides of \eqref{1.3} goes to $1$ as $\lambda\to\infty$ 
for any $b_n$ with $\sum\abs{b_n}<\infty$. 

Since
\begin{align*}
(E_n^\pm)^2-4 &= \abs{E_n^\pm \mp 2} \, \abs{E_n^\pm \pm 2} \\
&\geq 4\abs{E_n^\pm \mp 2},
\end{align*}
\eqref{1.3} implies that
\begin{equation} \label{1.4}
\sum_n \abs{E_n^+ -2}^{1/2} + \abs{E_n^- +2}^{1/2} \leq \f12 \biggl( \sum_n \abs{b_n} 
+ 4\sum_n \abs{a_n -1}\biggr).
\end{equation}
More generally, we will prove that

\begin{t2}
\begin{equation} \label{1.5}
\sum_n \abs{E_n^+ -2}^p + \abs{E_n^- +2}^p \leq c_p \biggl[\sum_n \abs{b_n}^{p+1/2} 
+4 \sum_n \abs{a_n-1}^{p+1/2}\biggr]
\end{equation}
for any $p\geq \f12$ where
\[
c_p = \f12 \, 3^{p-1/2} \,\f{\Gamma(p+1)}{\Gamma (p+\f32)}\, \f{\Gamma(2)}{\Gamma(\f32)}\, .
\]
\end{t2}

As for sums of moments for $p<\f12$, we will prove

\begin{t3} Let $0\leq p<\f12$. Let $\|\dott\|$ be any translation 
invariant norm on pairs of sequences $\{a_n\}_{n=0}^\infty$, $\{b_n\}_{n=0}^\infty$. For any 
$\veps >0$, there exists a Jacobi matrix with $a_n=1$, $b_n=0$ for $n$ large so that $\|(a,b)\|
\leq \veps$ but $\sum_n \abs{E_n^+ -2}^p + \abs{E_n^- -2}^p \geq \veps^{-1}$.
\end{t3}

As \eqref{1.3a} shows, \eqref{1.4} and \eqref{1.5} are poor as $\lambda\to\infty$, since the 
left side grows like $\lambda^p$ and the right side as $\lambda^{p+1/2}$. It is better to use 
\begin{align*}
(E_n^\pm)^2 -4 &= \abs{E_n^\pm -2} \, \abs{E_n^\pm \pm 2} \\
&\geq \abs{E_n^\pm -2}^2
\end{align*}
and \eqref{1.3} to obtain 
\begin{equation} \lb{1.6a} 
\sum_n \abs{E_n^+ -2} + \abs{E_n^- +2} \leq \sum_n \abs{b_n} + 4\sum_n \abs{a_n -1}
\end{equation} 
and the related

\begin{t4} 
\begin{equation} \lb{1.6b} 
\sum_n \abs{E_n^+ -2}^p + \abs{E_n^- +2}^p \leq \sum_n (b_n^+ + 2\abs{a_n -1})^p 
+ (b_n^- + 2\abs{a_n -1})^p
\end{equation}
\end{t4}

As \eqref{1.3a} shows, the ratio of the two sides of \eqref{1.6b} is $1$ as $\lambda 
\to\infty$. 

\smallskip

We got interested in this problem because Killip-Simon \cite{KS} needed a bound like 
Theorem~1 to prove a conjecture of Nevai \cite{PN1,PN2} that if the right side of 
\eqref{1.3} is finite, then a condition of Szeg\"o holds. They and we expected bounds like  
\eqref{1.3} to hold because of the analogous results for Schr\"odinger operators.

Nevai's conjecture says that if $\sum_n \abs{b_n} + \sum_n \abs{a_n -1}<\infty$, then, with 
$m$, the $m$-function defined by 
\[
m(E) = (J-E)_{11}^{-1}, 
\]
we have
\begin{equation} \label{1.5c}
\int_{-2}^2 \log \Ima m(E+i0)\, \f{dE}{\sqrt{4-E^2}}>-\infty.
\end{equation}

Killip-Simon \cite{KS} use a sum rule of Case \cite{C1, C2} that 
\begin{equation} \label{1.5a}
Z(m) = \-\sum_n \log \abs{a_n} + \sum \log \abs{\beta_j} 
\end{equation}
where $\beta_j$ is defined by $\abs{\beta_j}>1$ and $\beta_j + \beta_j^{-1}$ are the listing 
of the eigenvalues of $J$ outside $[-2,2]$. In \eqref{1.5a}, $Z(m)$ is defined by 
\begin{equation} \label{1.5b}
Z(m) = \f{1}{2\pi} \int_{-2}^2 \log \biggl( \f{\sqrt{4-E^2}}{\Ima m(E+i0)}\biggr) 
\f{dE}{\sqrt{4-E^2}}\,.
\end{equation}
\eqref{1.5a} is only proven initially for $J$ with $J-J_0$ finite rank. (Or, in any event, 
not initially for all $J$'s with $J-J_0$ trace class. Eventually, using our bounds here 
and the theory of Nevanlinna functions, Killip-Simon \cite{KS} do prove \eqref{1.5a} 
for trace class $J-J_0$.) Killip-Simon show $Z(m)$ is lower semicontinuous as a trace class 
$J$ is approximated by cutoff $J$'s with $J-J_0$ finite rank. Thus to prove $Z(m)<\infty$ 
(i.e., that \eqref{1.5c} holds), they need to control the right side of \eqref{1.5a}. Since 
$\sum \abs{a_n-1}<\infty$, the $\sum_n \log \abs{a_n}$ is absolutely convergent. Since 
$\abs{\beta_j}\sim 1+(\abs{E_j}-2)^{1/2}$ for $E_j$ close to $2$, \eqref{1.4} implies that 
$\sum \log \abs{\beta_j}$ is uniformly bounded.

Theorem~1 should also be interesting in connection with some recent results of 
Peherstorfer-Yuditskii \cite{PY}, who focus on the finiteness of the right side of 
\eqref{1.3}.

Bounds for Schr\"odinger operator eigenvalues of the form
\begin{equation} \label{1.6}
\sum_{n=1}^\infty \abs{E_n}^p \leq L_{p,\nu} \int_{\bbR^\nu} \abs{V(x)}^{p+\nu/2} \, 
d^\nu x 
\end{equation}
where $E_n$ are the negative eigenvalues of $-\Delta +V$ on $L^2 (\bbR^\nu)$ go back 
twenty-five years to the work of Lieb and Thirring \cite{LT1, LT2}, who used the case $p=1$, 
$\nu=3$ in their celebrated proof of the stability of matter. They proved \eqref{1.6} 
for $p>0$, $\nu\geq 2$, and $p>\f12$, $\nu=1$, and shortly thereafter, Cwikel \cite{Cw}, 
Lieb \cite{L1}, and Rozenblum \cite{Roz} proved \eqref{1.6} in case $p=0$, $\nu\geq 3$. 
It is easy to see (e.g.,  Landau-Lifshitz \cite[pp.~156--157]{LL} and Simon \cite{SiB}) that 
it is false in case $p=0$, $\nu=2$.

For many years, the case $p=\f12$, $\nu=1$ was open, perhaps in part because \cite{LT2} 
erroneously claimed to have proven it. Only in 1996 did Weidl \cite{We} establish this 
result for $p=\f12$, $\nu=1$. For $\nu=1$, Lieb-Thirring \cite{LT2} conjectured the 
optimal value of $L_{p,\nu}$ for all $p\geq \f12$. They proved their conjecture when 
$\nu=1$ for $p=\f32, \f52, \f72, \dots$, and subsequently, Aizenman-Lieb \cite{AL} for 
all $p\geq\f32$. Shortly after Weidl's work, Hundertmark, Lieb, and Thomas \cite{HLT} 
found a new proof which yielded the optimal constant $L_{1/2, 1}$. A partially alternate 
proof of a part of the argument in \cite{HLT} can be found in Hundertmark, Laptev, and 
Weidl \cite{HLW}.

Unlike the discrete case, the continuum theory has a scaling symmetry: taking $V(x)\to 
\lambda^2 V(\lambda x)$ yields $E_n\to \lambda^2 E_n$ since there is a unitary operator 
that implements $x\to\lambda x$. This forces the power $\abs{E}^p$ on the right side of 
\eqref{1.6} given the scaling behavior of $d^\nu x$. Thus the same power properly 
captures large and small $E$'s. In the discrete case, this is not so, which is why we 
have two bounds \eqref{1.4} and \eqref{1.6b}. As noted, \eqref{1.6b} is good for large 
coupling, but \eqref{1.4} is better for small $E$'s. In particular, if $b_n\sim n^{-\alpha}$ 
(with $\alpha >1$) for $n$ large, \eqref{1.6b} only implies $\sum\abs{E_n^+ -2}^p < 
\infty$ for $p>\alpha^{-1}$ which \eqref{1.4} implies is true for $p>\alpha^{-1} -\f12$.

Of course, the best extended estimate would involve powers of $(E^2 -4)^{1/2}$ but both 
the Aizenman-Lieb \cite{AL} method to increase powers and the Laptev-Weidl \cite{LW} method 
to increase dimension seem to require powers of $\text{dist}(E, \sigma_{\ess}(J))$. 
However, one can save a little bit of the structure; see the remark at the 
end of section \ref{S5}.

We note one interesting feature of \eqref{1.3} vis-\`a-vis the continuum bound. The continuum 
$p=\f12$ bound has an optimal constant, but is off by a factor of $2$ in the large coupling 
limit. For \eqref{1.3}, as we noted above, the optimal bound for small coupling is also 
exact in the large coupling limit. 

In Section~\ref{S2}, we will prove Theorem~1 when $a_n\equiv 1$ by closely following 
\cite{HLT} and then obtain Theorems~2 and 4 when $a_n\equiv 1$ by the now standard 
argument of Aizenman and Lieb \cite{AL}. In Section~\ref{S3}, we make a simple but 
useful observation that allows one to obtain estimates for eigenvalues for arbitrary 
Jacobi matrices from the estimates for the special case. Section~\ref{S4} contains some 
examples and some counterexamples, and proves Theorem~3. Section~\ref{S5} uses ideas of 
Laptev-Weidl \cite{LW} to prove bounds for the higher-dimensional case. In an appendix, 
we show how the ideas in this paper provide a simple proof of a strengthening of the 
Bargmann-type bound of Geronimo \cite{Ger1,Ger2}.

This paper is aimed towards two rather different audiences: the Schr\"odinger operator 
community and the orthogonal polynomial community, who have rather different toolkits. 
For that reason, we include some material (such as that at the start of Section~\ref{S2}) 
that one group or the other may regard as elementary.

\medskip
{\bf Acknowledgment.} We thank Jeff Geronimo, Fritz Gesztesy, Rowan Killip, and Paul Nevai for 
useful comments.

\bigskip
\section{Bounds for Discrete Schr\"odinger Operators}\label{S2}

In this section, we prove Theorems~1, 2, and 4 when all $a_n = 1$. We begin with some 
general preliminaries. Given any self-adjoint operator $A$, bounded from above, we 
define
\begin{equation} \label{2.1}
E_j^+ = \inf_{\varphi_1 \dots \varphi_{j-1}} \, \sup_{\substack{\psi :\psi\perp\varphi_j \\
\psi\in D,\, \|\psi\|=1}} \, \langle \psi, A\psi\rangle.
\end{equation}
Similarly, if $A$ is bounded below,
\begin{equation} \label{2.2}
E_j^- =\sup_{\varphi_i\dots\varphi_j} \, \inf_{\substack{\psi :\psi\perp\varphi_j \\
\psi\in D(A),\, \|\psi\|=1}}\, \langle \psi, A\psi\rangle
\end{equation}
We will use $E_j^\pm (A)$ if the dependence on $A$ is important. From the definitions, 
\begin{equation} \label{2.3}
A\leq B \Rightarrow E_j^\pm (A) \leq E_j^\pm (B)
\end{equation}
and
\begin{equation} \label{2.4}
E_1^- \leq E_2^- \leq \cdots \leq E_2^+ \leq E_1^+.
\end{equation}

The min-max principle (Theorem XIII.1 in Reed-Simon \cite{RS4}) asserts that
\begin{enumerate}
\item[(i)] $E_\infty^\pm =\lim E_j^\pm$ has $E_\infty^+(A)=\sup \sigma_{\ess}(A)$, $E_\infty^- 
(A) =\inf \sigma_{\ess}(A)$
\item[(ii)] If $A$ has $N^+$ (resp.~$N^-$) eigenvalues counting multiplicity in the interval  
$(E_\infty^+, \infty)$ (resp.~$(-\infty, E_\infty^-)$), these eigenvalues are precisely $E_1^\pm, 
E_2^\pm, \dots, E_{N_\pm}^\pm$ and $E_j^\pm = E_\infty^\pm$ if $j>N_\pm$.
\end{enumerate}

Next, note from the definition that if $A_m\to A$ in norm, then we have convergence of the 
corresponding eigenvalues since $\abs{E_j^\pm (A) - E_j^\pm (B)}\leq \|A-B\|$. It follows 
if $f$ is an arbitrary continuous nonnegative function, then
\begin{align*}
\sum_{j=1}^k f(E_j^\pm (A)) &=\lim_{m\to\infty}\, \sum_{j=1}^k f(E_j^\pm (A_m)) \\
&\leq \liminf_{m\to\infty} \, \sum_{j=1}^\infty f(E_j^\pm (A_m))
\end{align*}
so taking $k\to\infty$,
\begin{equation} \label{2.5}
\sum_{j=1}^\infty f(E_j^+ (A)) + f(E_j^- (A)) \leq \liminf_{m\to\infty}\, 
\sum_{j=1}^\infty f(E_j^+ (A_m)) + f(E_j^- (A_m)).
\end{equation}
\eqref{2.5} and the min-max principle imply

\begin{proposition}\lb{P2.1} To prove \eqref{1.3}--\eqref{1.5}, it suffices to prove the 
special case where only finitely many $a_n$'s differ from $1$ and finitely many $b$'s 
differ from $0$.
\end{proposition}

Next, we want to note the impact of restriction. Let $A$ be a bounded self-adjoint operator 
on $\calH$. Let $P$ be an orthogonal projection. By $A_P$, we mean $PAP$ restricted as an 
operator on $P\calH=\Ran P$. In \eqref{2.1}/\eqref{2.2}, changing from $A$ to $A_P$ adds 
the condition $\psi\in \Ran P$ and it decreases sups and increases infs. Thus

\begin{proposition}\lb{P2.2} 
\begin{equation} \label{2.6}
E_j^+(A_P)\leq E_j^+(A); \qquad E_j^- (A_P)\geq E_j^-(A)
\end{equation}
\end{proposition}

We have two applications of \eqref{2.6} in mind. First, given two two-sided sequences 
$\{a_n\}_{n=-\infty}^\infty, \{b_n\}_{n=-\infty}^\infty$, define the whole-line operator 
$W$ on $\ell^2 (\bbR)$ by
\begin{equation} \label{2.7}
(Wu)(n)=a_{n-1} u(n-1)+b_n u(n) + a_n u(n+1). 
\end{equation}
Thus, if $P$ is the projection of $\ell^2 (\bbZ)$ to $\ell^2(\bbZ_+)\subset \ell^2(\bbZ)$, 
$W_P=J$ where $J$ is built from the projected sequences $\{a_n\}_{n=1}^\infty$ and 
$\{b_n\}_{n=1}^\infty$. As a result, \eqref{2.6} implies

\begin{proposition}\lb{P2.3} To prove \eqref{1.3}--\eqref{1.5}, it suffices to prove the 
analogous result for the whole-line operators.
\end{proposition}

One might think that the results are much harder for whole-line operators. After all, it 
can be shown that if $\tilde b$ has compact support, then $J(a_n\equiv 1, b_n=\lambda
\tilde b_n)$ has no spectrum outside $[-2,2]$ if $\lambda$ is small, but $W(a_n\equiv 1, 
b_n=\lambda\tilde b_n)$ always has eigenvalues outside $[-2,2]$ if $\lambda\neq 0$, 
$\tilde b\not\equiv 0$. That is why there is a Bargmann bound for $J$ but not for $W$. 
However, it is {\it not} harder because \eqref{1.3}--\eqref{1.5} have translation invariant 
quantities for their right side. Let $P_n$ be the projection onto $\ell^2$ ($m\in\bbZ$, 
$m\geq n$). One can see that as $n\to -\infty$, $E_j^\pm (W_{P_n})\to E_j^\pm (W)$ so 
\eqref{1.3}--\eqref{1.5} for the Jacobi case actually implies it for the whole-line case.

The second application of \eqref{2.6} is to the study of the following objects that will 
play a role below:
\begin{equation} \label{2.8}
S_n^\pm (A) = \sum_{j=1}^n E_j^\pm (A). 
\end{equation}

\begin{proposition}\lb{P2.4} Let $A$ be a self-adjoint operator. 
\begin{enumerate}
\item[{\rm{(i)}}] $S_n^+ (A)=\sup\{\Tra(AP)\mid P^*=P, \, P^2=P, \, \Tra(P)=n\}$ 
\item[{\rm{(ii)}}] $S_n^- (A)=\inf \{\Tra(AP)\mid P^*=P,\, P^2=P,\, \Tra(P)=n\}$
\item[{\rm{(iii)}}] $A\mapsto S_n^+ (A)$ is convex; $A\mapsto S_n^- (A)$ is concave.
\end{enumerate}
\end{proposition}

{\it Remark.} One can see that if $E_\infty^+\geq 0$, then in (i) $P^2=P$ can be replaced by 
$\|P\|\leq 1$ which is how it is often written.

\begin{proof} (i) By \eqref{2.6}, $S_n^+ (A_P)\leq S_n^+(A)$. But since $\Ran P$ has 
dimension $n$, $S_n^+(A_P)=\Tra(A_P)=\Tra(AP)$. Thus
\[
S_n^+(A) \geq \sup\{\Tra(AP)\mid P^*=P, \, P^2 =P,\, \Tra (P)=n\}.
\]
Next, pick $\varphi_1, \dots,  \varphi_n$ as follows. If $n\leq N^+(A)$, take $\varphi_1, 
\dots, \varphi_n$ to be the eigenfunctions of $A$ with eigenvalues $E_1^+, \dots, E_n^+$. 
If $n>N^+$, pick $\varphi_1, \dots, \varphi_{N^+}$ to be the eigenfunctions for $A$ with 
eigenvalues $E_1^+, \dots, E_{N^+}^+$ and $\varphi_{N^+ +1}, \dots, \varphi_n$ to be arbitrary 
orthonormal vectors in $\Ran (P_{[E_\infty^+ -\veps, E_\infty^+]}(A))$, the range of 
the spectral projection which is infinite-dimensional when $N^+ <\infty$ since $E_\infty^+
=\sup \sigma_{\ess}(A)$.

Let $P$ be the projection onto the span of $\varphi_1, \dots, \varphi_n$. Then
\begin{align*}
\Tra(AP) &= \sum_{j=1}^n (\varphi_j, A\varphi_j) \\
&\geq S_n^+(A) -\veps [\min (n, N^+)-n].
\end{align*}
Since $\veps$ is arbitrary, 
\[
S_n^+(A) \leq \sup\{\Tra(AP)\mid P^*=P, \, P^2=P, \, \Tra(P)=n\}.
\]

\smallskip
(ii) The same proof as (i).

\smallskip
(iii) $S_n^\pm$ are the sup and inf of linear functions, so convex and concave, respectively. 
\end{proof}

As a final general preliminary, we note the Birman-Schwinger principle: Let $A$ be a self-adjoint 
operator which is bounded above with $\alpha=\sup \sigma(A)$. Let $B$ be a positive relatively form 
compact, that is,
\begin{equation} \label{2.8a}
K_\beta \equiv B^{1/2}(\beta-A)^{-1}B^{1/2}
\end{equation} 
is compact for one and hence for all $\beta >\alpha$. $K_\beta$ is called the Birman-Schwinger 
operator.

\begin{proposition}[The Birman-Schwinger Principle \cite{Bir, Sch}]\lb{P2.5} Let $\lambda> 0$. 
$\beta >\alpha$ is an eigenvalue of $A+\lambda B$ if and only if $K_\beta$ has eigenvalue 
$\lambda^{-1}$. We have for $j\leq N^+ (A+\lambda B)$,
\begin{equation} \label{2.9}
E_j^+ (K_{E_j^+ (A+\lambda B)})=\lambda^{-1}.
\end{equation}
\end{proposition}

{\it Remark.} The point of \eqref{2.9} is that the index $j$ is the same in both $E_j^+$'s.

\begin{proof} For simplicity, we suppose $A$ and $B$ are bounded operators,  which is 
true in the applications we will make. 
If $(A+\lambda B)\varphi=\beta\varphi$, then $B^{1/2}(\beta - A)^{-1} B^{1/2} 
(B^{1/2}\varphi)=\lambda^{-1} B^{1/2}\varphi$ and $B^{1/2}\varphi\neq 0$ since if not, we 
must have $A\varphi=\beta\varphi$, which is impossible since $\beta >\sup \sigma(A)$. 
Conversely, if $K_\beta \psi =\lambda^{-1}\psi$ and $\varphi=(\beta -A)^{-1}B^{1/2}\psi$ 
($\neq 0$ since $\lambda^{-1}\neq 0$), we have $(A+\lambda B)\varphi=\beta\varphi$. Thus 
the first expression is true.

Next, note that $\|K_\beta\|\to 0$ as $\beta\to\infty$ by compactness. Its eigenvalues 
are continuous, and so by eigenvalue perturbation theory \cite{Kato, RS4}, real analytic. 
If $e(\beta)$ is a positive eigenvalue of $K_\beta$ with $K_\beta \varphi =e\varphi$ and 
$\|\varphi\|=1$, then by eigenvalue perturbation theory (the Feynman-Hellmann theorem), 
\[
\f{de}{d\beta}=\biggl\langle \varphi, \f{\partial K_\beta}{\partial \beta}\varphi
\biggr\rangle = -\|(\beta -A)^{-1} B^{1/2} \varphi\|^2 <0,
\]
so $e$ is strictly monotone. Thus if $e(\beta)$ is the $j$th eigenvalue of $K_\beta$ and   
$e(\beta_0)>\lambda^{-1}$, there is exactly one 
$\beta >\beta_0$ with $e(\beta)=\lambda^{-1}$, so
\[
\# \{j\mid E_j^+ (K_{\beta_0})\geq \lambda^{-1}\}=\#\{\beta >\beta_0\mid E_j^+ 
(K_\beta)=\lambda^{-1}\}
\]
(counting multiplicity) from which \eqref{2.9} follows.
\end{proof}

With the general preliminaries out of the way, we compute the Birman-Schwinger operator for 
$A=W_0$ and  a diagonal (i.e., $a_n\equiv 1$) perturbation.

\begin{proposition} \lb{P2.6} Let $W_0$ be the whole-line matrix with $a_n\equiv 1$, $b_n 
\equiv 0$. Let $\beta >2=\sup \sigma(W_0)$. Then $(\beta -W_0)^{-1}$ has matrix elements
\begin{equation} \label{2.10}
[(\beta - W_0)^{-1}]_{nm} = (\mu^{-1} -\mu)^{-1} \mu^{\abs{n-m}} 
\end{equation}
where $\mu$ is related to $\beta$ by
\begin{equation} \label{2.11}
\beta =\mu+\mu^{-1}; \qquad \mu <1.
\end{equation}
\end{proposition}

{\it Remark.} Of course, $\mu=\f12 (\beta -\sqrt{\beta^2-4}\,)$ and $\mu^{-1}=\f12 
(\beta + \sqrt{\beta^2-4}\,)$ so $\mu^{-1}-\mu =\sqrt{\beta^2 -4}$. This is why  
$\sqrt{E^2 -4}$ enters in Theorem~1.

\begin{proof} This is a standard calculation. Looking for solutions of
\begin{equation} \label{2.12}
\varphi (n-1) + \varphi(n+1) =\beta\varphi(n),
\end{equation}
one tries $\varphi(n)=\zeta^n$ and finds $\zeta + \zeta^{-1}=\beta$, so the solutions are 
$\zeta=\mu$ and $\zeta=\mu^{-1}$. Let
\[
\varphi_\pm (n)=\mu^{\pm n}.
\]
Both solve \eqref{2.12} if $\mu$ obeys \eqref{2.11}. Since $\mu <1$, $\varphi_+$ is $\ell^2$ 
at $+\infty$, $\varphi_-$ at $-\infty$, so the right side of \eqref{2.10} which has the 
form $(\mu^{-1} -\mu)^{-1}\varphi_- (\min(n,m))\varphi_+ (\max(n,m))\equiv G_n(m)$ is 
$\ell^2$ in $m$ for each $n$ with $((W_0-\beta)G_n)(m)=0$ if $m\neq n$. By a direct 
computation (essentially $\mu^{-1} -\mu$ is the Wronskian of $\varphi_+$ and $\varphi_-$), 
$(\beta_0 -W)G_n=\delta_n$, that is, $G_n(m)=[(\beta_0 -W)^{-1}\delta_n](m)$, proving 
\eqref{2.10}. 
\end{proof}

{\it Remark.} Alternatively, one can use Fourier analysis to compute the inverse.

\smallskip
Because of \eqref{2.10}, the following operator will enter in our discussion, 
$\{b_n\}_{n\in\bbZ}$ is a positive sequence of finite support,
\begin{equation} \label{2.13}
(L_\mu)_{nm} = b_n^{1/2} \mu^{\abs{n-m}} b_m^{1/2}.
\end{equation}
Recall the definition \eqref{2.8} of $S_m (\dott)$. The crucial lemma is

\begin{proposition}\lb{P2.7} Let $0<\mu<\eta\leq 1$. Then for any $n$,
\begin{equation} \label{2.14}
S_n^+ (L_\mu) \leq S_n^+ (L_\eta).
\end{equation}
\end{proposition}

{\it Remarks.} 1. Since $\Tra (L_\mu)$ is constant, individual eigenvalues cannot all 
be monotone.

\smallskip
2. This is a special case of the warm-up to the proof of Lemma~4 in \cite{HLT}. 
Our proof is close to the proof there, except where \cite{HLT} uses eigenvalue perturbation 
at $\mu_j=0$, we use symmetry.

\begin{proof} Given a bounded positive sequence $\{\mu_n\}_{n=-\infty}^\infty$, we define 
\begin{alignat*}{2}
(L_{\{\mu_n\}})_{k\ell} &= b_k^{1/2} b_\ell^{1/2} \prod_{j=k}^{\ell-1}\mu_j 
\qquad && \text{if } k\leq \ell \\
&= (L_{\{\mu_n\}})_{k\ell} \qquad &&\text{if } k>\ell
\end{alignat*}
so $L_\mu$ is $L_{\{\mu_n\}}$ when all $\mu_n=\mu$. Thus \eqref{2.14} follows if we show 
$S_n^+(L_{\{\mu_n\}})$ is monotone in $\mu_n\in [0,\infty)$ when $\{\mu_j\}_{j\neq n}$ are 
held fixed. Let $f(\mu)$ be this function when $\mu_n$ takes the value $\mu$. 
$L_{\{\mu_j\}_{j\neq n},\, \mu_n =\mu}$ is affine in $\mu$ for each matrix element is 
either constant or a multiple of $\mu$. More precisely,  in matrix notation we have 
\begin{displaymath}
L_{\{\mu_j\}_{j\neq n}, \mu_n =\mu}= 
\begin{pmatrix} 
A& 0\\
0 & B 
\end{pmatrix} +
\mu 
\begin{pmatrix} 
0& C\\
C^\dag & 0 
\end{pmatrix} 
\end{displaymath}
where $A,B$, and $C$ depend only on $\{\mu_j\}_{j\not = n}$. So by Proposition~\ref{P2.4} (iii), 
$f(\mu)$ is a convex function of $\mu$.

On the other hand, if $U$ is the diagonal matrix,
\begin{alignat*}{2}
(U\varphi)(\ell) &=-\varphi(\ell) \qquad &&\ell\leq n \\
&=\varphi(\ell) \qquad &&\ell \geq n+1,
\end{alignat*}
or, as a block matrix, $U= \left (\begin{smallmatrix} 1 & 0\\0& -1\end{smallmatrix}\right)$, 
then $UL_{\{\mu\}}U^{-1}=L_{\{\tilde\mu\}}$ where
\begin{alignat*}{2}
\tilde\mu_\ell &= \mu_\ell \qquad &&\text{if }\ell\neq n \\
&=-\mu_\ell \qquad &&\text{if }\ell=n ,
\end{alignat*}
that is, we have
\begin{displaymath}
U\begin{pmatrix}
A & \mu C \\
\mu C^\dag & B
\end{pmatrix}
U^{-1}
= 
\begin{pmatrix}
A & -\mu C \\
-\mu C^\dag & B
\end{pmatrix}
= L_{\{\mu_j\}_{j\neq n},\, \mu_n =-\mu}. 
\end{displaymath}
Since $E_j^+$, and so $S_j^+$, are invariant under unitary transformations, we see 
$f(-\mu)=f(\mu)$. An even convex function is monotone increasing on $[0,\infty)$, so 
$S_n^+ (L_{\{\mu_n\}})$ is monotone in each $\mu_n$ in the region $\mu_n \geq 0$. 
\end{proof}

We are now ready to prove what is essentially Theorem~1 in case $a_n\equiv 1$.

\begin{theorem} \lb{T2.8} Let $W_0$ be the free whole-line Schr\"odinger operator and 
$B$ a positive finite-rank diagonal matrix. Let $W=W_0 +B$. Then
\begin{equation} \label{2.14a}
\sum_{j=1}^{N_+(W)} \sqrt{E_j^+ (W)^2-4} \leq \Tra(B).
\end{equation}
\end{theorem}

\begin{proof} (Following \cite{HLT}) Since $B$ is finite rank, we know that $N_j^+(W)<\infty$. 
Define $\mu_j$ by $\mu_j^{-1} + \mu_j =E_j^+$ with $\mu_j<1$. By \eqref{2.8a} and the 
remark after Proposition~\ref{P2.6},
\begin{equation} \label{2.15}
K_{E_j^+}=((E_j^+)^2 -4)^{-1/2} L_{\mu_j}
\end{equation}
with $L_\mu$ given by \eqref{2.13}. By \eqref{2.9}, 
\begin{equation} \label{2.16}
E_j^+ (K_{E_j^+})=1.
\end{equation}
Since for $a>0$, $E_j^+ (aA) =aE_j^+(A)$, \eqref{2.15}, \eqref{2.16} imply
\begin{equation} \label{2.17}
\sqrt{E_j^+ (W)^2 -4} = E_j^+ (L_{\mu_j}).
\end{equation}
Thus
\begin{equation} \lb{2.new}
\sum_{j=1}^{N^+(W)}\sqrt{E_j^+ (W)^2 -4} = 
E_1^+ (L_{\mu_1}) + E_2^+ (L_{\mu_2}) + \cdots + E_{N^+}^+ (L_{\mu_{N^+}}).
\end{equation}
But, by \eqref{2.14} and $\mu_1 < \mu_2 < \cdots < \mu_{N_+}<1$,
\begin{align*}
E_1^+ (L_{\mu_1}) + E_2^+ (L_{\mu_2}) &= S_1^+ (L_{\mu_1}) + E_2^+ (L_{\mu_2}) \\
&\leq S_1^+ (L_{\mu_2}) + E_2^+ (L_\mu) \\
&=S_2^+ (L_{\mu_2}) \\
&\leq S_2^+ (L_{\mu_3}),
\end{align*}
so by induction,
\[
\sum_{j=1}^k E_j^+ (L_{\mu_j}) \leq S_k^+ (L_{\mu_k}) \leq S_k^+ (L_{\mu_{k+1}})
\]
and thus \eqref{2.new} implies
\[
\sum_{j=1}^{N^+(W)} \sqrt{E_j^+ (W)^2 -4} \leq \sum_{j=1}^{N^+} E_j^+ (L_{\mu=1})=
\Tra(B)
\]
since $L_{\mu=1}$ is the rank one operator $b_n^{1/2} b_m^{1/2}$ with a single nonzero 
eigenvalue equal to $\Tra(L_{\mu=1})=\Tra(B)$.
\end{proof}

{\it Remark.} The proof shows the inequality is strict if $E_1^+ (L_\mu)$ is strictly 
monotone.  Thus the inequality is strict if rank $(B)\geq 2$.

\smallskip
There is a standard argument of Aizenman-Lieb \cite{AL} which we can use to go from a 
$(\f12, 1)$ bound (power of $E-2$, power of $b$) to a general $(p, p+\f12)$ for any 
$p\geq \f12$:

\begin{theorem} \lb{T2.9} Under the hypothesis of Theorem~\ref{T2.8}, we have, for any 
$p\geq\f12$,
\begin{equation} \label{2.18}
\sum_{j=1}^{N_+(W)} \abs{E_j^+(W)-2}^p \leq \f12 \, \f{\Gamma(p+1)}{\Gamma (p+\f32)} 
\, \f{\Gamma (2)}{\Gamma(\f32)}\, \Tra(\abs{B}^{p+1/2}).
\end{equation}
\end{theorem}

\begin{proof} Note first that since $b_n \leq (b_n)_+\equiv \max (0, b_n)$, if positivity of 
$B$ is dropped, we still have that
\begin{equation} \label{2.19}
\sum_{j=1}^{N_+(W)} \abs{E_j^+(W)-2}^{1/2} \leq \f12\sum_n (b_n)_+
\end{equation}
by using \eqref{2.5}, $W_0 + B\leq W_0 + B_+$ and $\sqrt{E^2 -4}\leq 2\abs{E-2}^{1/2}$.

Let $r>0$. Then
\[
(E_j^+(W)-2-r)_+ = (E_j^+(W-r\boldsymbol{1})-2)_+
\]
so \eqref{2.19} implies
\begin{equation} \label{2.20}
\sum_{j=1}^{N_+(W)} \abs{E_j^+ (W) -2-r}_+^{1/2} \leq \f12 \sum_n (b_n -r)_+.
\end{equation}

Now the well-known integral for $\alpha <p$, 
\[
\f{\Gamma (p+1)}{\Gamma (p-\alpha)\Gamma (\alpha+1)} \int_0^1 (1-x)^\alpha x^{p-\alpha -1}\, 
dx =1
\]
with scaling implies for any $\alpha <p$:
\begin{equation} \label{2.21}
a_+^p = C_{p,\alpha} \int_0^\infty (a-r)_+^\alpha r^{p-\alpha-1}\, dr 
\end{equation}
where $C_{p,\alpha} =\Gamma(p+1)/\Gamma(p-\alpha)\Gamma(\alpha+1)$. \eqref{2.20} and 
\eqref{2.21} immediately imply that
\begin{equation} \lb{2.24b}
\sum_{j=1}^{N_+(W)} \abs{E_j^+ (W) -2}^p \leq \f12 \,\f{C_{p,1/2}}{C_{p+1/2, 1}} \, 
\sum_n (b_n)_+^{p+1/2}
\end{equation}
which implies \eqref{2.18}.
\end{proof}

Similarly, we have

\begin{theorem} \lb{T2.10} Under the hypothesis of Theorem~\ref{T2.8}, we have for any 
$p\geq 1$,
\begin{equation} \lb{2.24a} 
\sum_{j=1}^{N_+ (W)} \abs{E_j (W) -2}^p \leq\Tra (\abs{B}^p).
\end{equation} 
\end{theorem}

\begin{proof} Since $E\geq 2$ implies 
\begin{align*}
E^2 -4 &= (E-2)^2 (E+2) \\
&\geq (E-2)^2,
\end{align*}
\eqref{2.14a} implies \eqref{2.24a} for $p=1$. The result for general $p\geq 1$ follows 
as above. Where above we get a factor of $C_{p,1/2}/C_{p-1/2, 1}$, here we get 
$C_{p,1}/C_{p,1} =1$. 
\end{proof}

So far, we have proven a bound on $E_j^+$, but they immediately imply bounds on $E_j^-$. One 
can prove that by analogy, but it is even easier to use the unitary map
\[
(Vu)(n)=(-1)^n u(n)
\]
which has
\[
VW(\{a_n\}, \{b_n\})V^{-1} =W(\{-a_n\}, \{b_n\}) = -W (\{a_n\}, \{-b_n\}), 
\]
so
\begin{equation} \label{2.22}
E_j^- (W(\{a_n\}, \{b_n\}))= -E_j^+ (W(\{a_n\}, \{-b_n\})).
\end{equation}
Thus, for example,
\begin{align} 
\sum_{j=1}^{N_r} \abs{E_j^-(W)^2 -4}^{1/2} &\leq \sum_n (-b_n)_+ \notag \\
&= \sum_n (b_n)_- \lb{2.23}
\end{align}
where $x_- =(-x)_+ = -\min(0,x)$ so $\abs{x}=x_+ + x_-$ and we obtain \eqref{1.3} for 
the case $a_n\equiv 1$.

\bigskip
\section{Bounds for Jacobi Matrices}\label{S3}

The following elementary observation lets us pass from bounds in case $a_n\equiv 1$ to the 
general case. Note that
\[
\begin{pmatrix} -\abs{a_n-1} & 1 \\ 1 & -\abs{a_n-1} \end{pmatrix} 
\leq 
\begin{pmatrix} 0 & a_n \\ a_n & 0 \end{pmatrix}
\leq 
\begin{pmatrix} \abs{a_n-1} & 1 \\ 1 & \abs{a_n -1} \end{pmatrix}
\]
for any $a_n$ real since for any $x$ in $\bbR$, $\left( \begin{smallmatrix} \abs{x} & x \\
x & \abs{x} \end{smallmatrix}\right) \geq 0$ since it has determinant $0$ and trace 
$2\abs{x}\geq 0$. This immediately implies by repeated use at each pair of indices 
\begin{equation} \label{3.1}
W(\{a_n \equiv 1\}, \{b_n^-\}) \leq W(\{a_n\}, \{b_n\}) \leq W(\{a_n\equiv 1\}, \{b_n^+\})
\end{equation} 
where
\begin{equation} \label{3.2}
b_n^\pm = b_n \pm (\abs{a_{n-1} -1} + \abs{a_n -1}).
\end{equation}
\eqref{3.1} and \eqref{2.5} immediately imply

\begin{theorem}\lb{T3.1} Let $f$ be monotone increasing on $(0,\infty)$ and even. Then
\begin{equation} \label{3.3}
f(E_j^\pm (W(\{a_n\}, \{b_n\}))) \leq f(E_j^\pm (W(\{a_n\equiv 1\}, \{b_n^\pm\}))) 
\end{equation}
where $b_n^\pm$ is given by \eqref{3.2}.
\end{theorem}

With this, we can now prove our three main theorems:

\begin{proof}[Proof of Theorem~1] By \eqref{3.3}, \eqref{2.20}, and \eqref{2.23}, 
\begin{align}
\sum_n & [(E_n^+)^2 -4]^{1/2}  + [(E_n^-)^2 -4]^{1/2} \notag \\
&\leq \sum_n [b_n + \abs{a_{n-1}-1} + \abs{a_n-1}]_+ + 
[b_n -\abs{a_{n-1}-1} - \abs{a_n -1}]_-  \notag \\
&\leq \sum_n [b_n]_+ + [b_n]_- + 4\sum_n \abs{a_n-1}. \lb{3.4}
\end{align}
In obtaining \eqref{3.4}, we used $[x+y]_+ \leq x_+ + y_+$ and $[x+y]_- \leq x_- + y_-$ and 
that a given $\abs{a_n -1}$ occurs in four terms with $[b_n]_\pm$ and $[b_{n+1}]_\pm$. 
\end{proof}

\begin{proof}[Proof of Theorem~2] By \eqref{3.3}, \eqref{2.18}, and \eqref{2.22}, 
\begin{equation} \label{3.5}
\begin{split}
\sum_n & \abs{E_n^+ -2}^p  + \sum_n \abs{E_n^- +2}^p \leq \\
& \quad d_p \sum_n [b_n + \abs{a_{n-1}-1} 
+ \abs{a_n -1}]_+^{p+1/2} + [b_n - \abs{a_{n-1} -1} - \abs{a_n -1}]_-^{p+1/2} 
\end{split}
\end{equation}
where
\[
d_p = \f12 \, \f{\Gamma(p+1)}{\Gamma(p+\f32)} \, \f{\Gamma(2)}{\Gamma (\f32)}\,.
\]

Now for any $q\geq 1$ ($q$ will be $p+\f12$), $x^q$ is convex, so
\begin{align*}
(\alpha + \beta + \gamma)^q &= 3^q \biggl( \f{\alpha}{3} + \f{\beta}{3} + 
\f{\gamma}{3}\biggr)^q \\
&\leq 3^{q-1} [\alpha^q + \beta^q + \gamma^q]
\end{align*}
from which \eqref{1.5} holds if we note that $c_p =3^{(p+1/2)-1}\, d_p$. 
\end{proof}

\begin{proof}[Proof of Theorem~4] As stated, \eqref{1.6b} is an immediate consequence of 
Theorem~\ref{T2.10} and \eqref{3.1}. 

We kept this bound in the form \eqref{1.6b} to get an exact result as $\lambda\to\infty$. 
We could use the same method of proof of Theorem~2 to get 
\[
\sum_n \abs{E_n^+ -2}^p + \sum_n \abs{E_n^- +2}^p \leq 3^{p-1} \biggl[ \sum_n \abs{b_n}^p 
+ 4\sum_n \abs{a_n -1}^p \biggr].
\]
\end{proof}

We believe it could be true that \eqref{1.3} holds with $\abs{a_n -1}$ replaced by 
$(a_n -1)_+$ and, in particular, we know that \eqref{1.5} and \eqref{1.6b} hold when 
$p\geq 1$ if $\abs{a_n -1}$ is replaced by $(a_n -1)_+$. To see the latter, we note 
that --- by a convexity plus evenness argument much like that in the proof of 
Proposition~\ref{P2.7} --- $\sum_{j=1}^k E_j^+ (W(\{a_n\}, \{b_n\}))$ is monotone 
in $a_n$ in the region $a_n \geq 0$. Thus for $p=1$, \eqref{1.5} and \eqref{1.6b} hold 
with $(a_n -1)_+$ for we move those $a$'s with $a_n >1$ to the diagonal as we did in 
\eqref{3.1}, and use the monotonicity just noted to move $a_n$'s in $(0,1)$ up to $1$. 
Once one has the result for $p=1$, it follows for $p\geq 1$ by the Aizenman-Lieb argument.

The fact that in \eqref{1.5} for $p\geq 1$ and in the Bargmann bound of the Appendix, one 
can take $(a_n -1)_+$ leads us to conjecture \eqref{1.3} holds with $(a_n -1)_+$ 
rather than $\abs{a_n -1}$.

\bigskip
\section{Examples}\label{S4}

\begin{example} \lb{E4.1} $W$ has all $a_n =1$, all $n$, and all $b_n =0$ for $n\neq 0$. 
If $b_0\equiv b >0$, then there is an eigenvalue at energy $E=\mu + \mu^{-1}$ with $\mu <1$ 
and eigenfunction $\varphi_n = \mu^{\abs{n}}$. To have the eigenfunction fit at $n=0$, we 
need 
\[
2\mu + b1=E1
\]
or
\[
b=\mu^{-1} -\mu = \sqrt{E^2-4}\, .
\]
This example has equality in \eqref{1.3} for all values of $b>0$ (and also $b<0$ it turns out) 
and shows one cannot decrease the value $1$ in front on $\sum\abs{b_n}$.
\qed
\end{example}

\begin{example} \lb{E4.2} $W$ has all $b_n=0$, all $n$, and all $a_n =1$, $n\neq 0$. If 
$a_0 \equiv a >1$, there is an eigenvalue at energy $E=\mu + \mu^{-1}$ with $0<\mu <1$. 
Then $\varphi_n =\mu^{-n}$ for $n\leq 0$ and $\varphi_n =\mu^{n-1}$ for $n\geq 1$ 
since $\varphi$ must be symmetric around $n= \f12$. The eigenfunction condition at $0$ reads 
\[
\mu+a = \mu + \mu^{-1}
\]
or $a=\mu^{-1}$. Thus
\[
a - a^{-1} = \sqrt{E^2-4}\, .
\]
There is a second eigenvalue at energy $-E$ (there has to be by the symmetry \eqref{2.22}). 
Thus 
\begin{align*}
\text{LHS of \eqref{1.3}}&= 2(a-a^{-1}) \\
&= 2(1+a^{-1})(a-1).
\end{align*}
The two sides of \eqref{1.3} are not equal for any $a$, but the ratio goes to $1$ as 
$a\downarrow 1$ since $2(1+a^{-1})\uparrow 4$. Thus the $4$ in front of the $\abs{a-1}$ cannot 
be made smaller. However, both this example and the discussion in the appendix suggest it might 
be possible to replace $\abs{a-1}$ by $(a-1)_+$. \qed
\end{example}

As noted above, the best constant for the $W$ case is the same as for the $J$ case.

\begin{example}[Proof of Theorem~3]\lb{E4.3} Shift to the Jacobi case. Take an example 
with $a_n \equiv 1$ and $b_n=0$, except for $n=m, 2m, \dots, Nm$ where $b_n =\beta$. As 
$m\to\infty$, there are $n$ eigenvalues above $2$ which all approach the solution of 
$\sqrt{E^2 -4} =\beta$. So long as $\beta<1$, $\abs{E_n-2} \geq \f16\beta^2$, so
\begin{equation} \label{4.1}
\sum_n \abs{E_n -2}^p \geq N\biggl(\f{\beta^2}{6}\biggr)^p.
\end{equation}

In the translation invariant norm $\|\dott\|$, let $\alpha =\|(a_n \equiv 1, b_1 =1, 
b_n =0\text{ for } n\neq 1)\|$. Then for the $(a,b)$ of this $\beta,N,m$ example, 
\begin{equation} \label{4.2}
\|(a,b)\|\leq N\alpha\beta.
\end{equation}

Let $N_0(\veps)$, $\beta_0(\veps)$ solve
\begin{align*}
N\biggl(\f{\beta}{6}\biggr)^p &= 2\veps^{-1} \\
N\alpha\beta &= \f{\veps}{2}
\end{align*}
so
\begin{align*}
\beta &= c_1 \veps^{2/1-2p}\to 0 \\
N &= c_2 \veps^{-(1+2p)/(1-2p)} \to \infty
\end{align*}
since $p<\f12$. Increase $N$ slightly to be an integer. Thus 
\[
\sum_n \abs{E_n -2}^p \geq \veps^{-1}, \qquad \|(a,b)\|\leq \veps,
\]
proving Theorem~3. 
\qed
\end{example}

\bigskip
\section{Bounds in Higher Dimension} \label{S5}

In this section, we want to use the ideas of Laptev-Weidl \cite{LW} to prove bounds on 
operators on $\ell^2 (\bbZ^\nu)$. We begin with the discrete Schr\"odinger operator case. 
Let $H_0$ be defined on $\ell^2 (\bbZ^\nu)$ by 
\[
(H_0 u)(n) = \sum_{\abs{m-n}=1} u(m)
\]
and
\[
(Vu)(n) = V(n)u(n).
\]

\begin{lemma} \lb{L5.1} Let $W_0$ act on $\ell^2 (\bbZ;X)$ where $X$ is a Hilbert space, 
and let $B(n):X\to X$ be self-adjoint and trace class with $\sum_n \Tra (\abs{B(n)})<\infty$. 
Then 
\begin{equation} \label{5.1}
\sum_j (E_j^\pm (W_0 + B)^2 -4)^{1/2} \leq \Tra_X (B^\pm) 
\end{equation}
where $B^\pm (n) =\max(\pm B(n),0)$ is defined via the functional calculus.
\end{lemma}

\begin{proof} Suppose $B(n) \geq 0$. As with \eqref{2.13}, define $L_\mu :\ell^2 (\bbZ;X) 
\to \ell^2 (\bbZ;X)$ by 
\[
(L_\mu)_{mn} = B_n^{1/2} \mu^{\abs{n-m}} B_m^{1/2}.
\]
As with Proposition~\ref{P2.7}, $0< \mu <\eta \leq 1$ implies 
\[S_n^+ (L_\mu) \leq S_n^+ (L_\eta)
\]
and then the proof of \eqref{2.14a} extends. 
\end{proof}

\begin{theorem} \lb{T5.2}  If $V\in L^p (\bbZ^\nu; X)$ for $p\geq 1$ where $X$ is a Hilbert space, 
that is, $V(x):\, X\to X$ is a symmetric compact operator such that $\sum_{x\in\bbZ^\nu} 
\Tra_X\abs{V(x)}^p<\infty$, then 
\begin{equation} \label{5.2}
\sum_j \abs{E_j^+ (H_0 +V) -2\nu}^p +\sum_j \abs{E_j^- (H_0 +V) + 2\nu}^p \leq \sum_{x\in \bbZ^\nu} 
\Tra_X \abs{V(x)}^p.
\end{equation}
\end{theorem}

\begin{proof} By the Aizenman-Lieb idea, \eqref{2.21}, it suffices to prove this for $p=1$. 
As usual, we can suppose $V\geq 0$ and prove the result for $E_j^+$. Write 
\[
H_0 = H_{0,1} + H_{0,\{2,\dots,\nu\}} 
\]
where $H_{0,1}$ involves neighbors in the $1$ direction and $H_{0,\{2, \dots,\nu\}}$ 
neighbors in the other directions. Note that  
\begin{equation} \label{5.3}
(H_{0,1} + H_{0,\{2,\dots,\nu\}}+V-2\nu)_+ \leq (H_{0,1} + (H_{0,\{2,\dots,\nu\}} 
+V-2(\nu -1))_+ -2)_+
\end{equation}
and thus  
\begin{align*}
\sum_j \abs{E_j^+ &(H_0 +V) -2\nu} = \Tra_{\ell^2(\bbZ^\nu; X)} ((H_0 + V-2\nu)_+) \\
&\leq \Tra_{\ell^2(\bbZ; \ell^2(\bbZ^{\nu-1};X))} ((H_{0,1} + (H_{0,\{2,\dots,\nu\}} + 
V-2(\nu -1))_+ -2)_+) \\
&\leq \sum_{n_1} \Tra_{\ell^2(\bbZ^{\nu-1};X)} ((H_{0,\{2, \cdots, \nu\}} + 
V(n_1, \dott) -2\nu +2)_+) 
\end{align*}
by \eqref{5.1} and $(E^2 -4)^{1/2} \geq (\abs{E}-2)$. An inductive argument completes the 
proof. 
\end{proof}

For the other moment result, it will be convenient to phrase things in terms of the 
classical constants, 
\begin{align} 
L_{p,\nu}^{c\ell} &= (2\pi)^{-\nu/2} \int_{\abs{k}\leq 1} \abs{k}^{2p}\,  d^\nu k 
\notag \\
&= 2^{-\nu} \pi^{-\nu/2} \,\f{\Gamma (p+1)}{\Gamma (p+1 + \f{\nu}{2})}\,. \label{5.4}
\end{align}
These constants have several important features. First, the argument that led to \eqref{2.24b} 
says that if 
\begin{equation} \label{5.5}
\sum_{j=1}^{N_+} \abs{E_j^+ -2}^p \leq \alpha L_{p,\nu}^{c\ell} \sum_n \abs{b_n}^{p+\nu/2}
\end{equation}
for some $p=p_0$, it holds for all $p>p_0$. Second, 
\[
L_{p=1/2,\,\nu=1} =2^{-1} \pi^{-1/2} \biggl[ \f{\f12\sqrt{\pi}}{1}\biggr] =\f14
\]
so the consequence of \eqref{5.1} and $(E^2 -4)^{1/2} \geq 2(\abs{E}+2)^{1/2}$ is that 
\eqref{5.5} holds for $\nu =1$, $p=\f12$, and $\alpha=2$. 

Finally, we note that from \eqref{5.4} and Fubini, we have 
\begin{equation} \label{5.6}
L_{p,\nu}^{c\ell} = \prod_{j=0}^{\nu=1} L_{p+j/2, 1}^{c\ell}.
\end{equation}

\begin{theorem} \lb{T5.3} Let $V\in L^{p+\nu/2} (\bbZ^\nu; X)$ for $p\geq 1$. Then  
\begin{equation} \label{5.7}
\sum_j \abs{E_j^+ (H_0 +V) -2\nu}^p + \sum_j \abs{E_j^- (H_0 +V)+2\nu}^p \leq 
2^\nu L_{p,\nu}^{c\ell} \sum_{x\in \bbZ^\nu} \Tra_X \abs{V(x)}^{p+\nu/2}.
\end{equation}
\end{theorem}

\begin{proof} We exploit \eqref{5.3}, but use \eqref{5.5} for $\alpha=2$, $\nu=1$, 
$p\geq \f12$ at each stage of the induction. We then get \eqref{5.7} with a constant 
\[
\prod_{j=0}^{\nu-1} 2L_{p+1/2, 1}^{c\ell} =2^\nu L_{p,\nu}^{c\ell}
\]
by \eqref{5.6}. 
\end{proof}

As in the one-dimensional case, Theorem~\ref{T5.2} is better for large coupling. Indeed, 
it is exact in the large coupling regime, while Theorem~\ref{T5.3} gives more information 
on the eigenvalues very close to $\pm 2\nu$ in the regime of slow decay of $V(n)$ at infinity.

As with the one-dimensional case, we can handle nonconstant off-diagonal terms which approach 
$1$ fast enough at infinity. Let $B(\bbZ^\nu)$ be set of bonds in $\bbZ^\nu$, that is, the set 
of unordered pairs  $b=(ij)$ with $i,j\in\bbZ^\nu, \abs{i-j}=1$. Given 
$\{a_b\}_{b\in B(\bbZ^\nu)}$, a nonnegative real number $a_b$ for each bond $b=(ij)$, one 
can define 
\begin{equation} \label{5.8}
(H_0 (a_b)u) (n) = \sum_{\abs{m-n} =1} a_{(nm)} u(m).
\end{equation}
The analog of \eqref{3.3} is then 
\begin{equation} \label{5.9}
H_0 +V^- \leq H_0 (a_b) +V \leq H_0 +V^+
\end{equation}
where
\begin{equation} \label{5.10}
V^\pm (n) = V(n) \pm \sum_{\abs{m-n} =1} \abs{a_{(mn)} -1}
\end{equation}
so, for example, we get 
\begin{equation} \label{5.11}
\sum_j [E_j^+ (H_0 (a_b) +V) + E_j^- (H_0 (a_b)+V)] \leq \sum_n \abs{V(n)} + 
\sum_b 4\abs{a_b-1}.
\end{equation}

{\it Remark.} Since the bound in Theorem~1 is optimal both for large and small coupling, 
the curious reader might wonder whether it is possible to keep some of its structure also in 
higher dimension. This is indeed the case. For constant diagonal terms and scalar potential 
we have the two bounds
\begin{displaymath}
\sum_{n=1,\dots, N_\pm} [(E_n^+)^2 -4]^{1/2} + [(E_n^-)^2 -4]^{1/2} \leq 
\sum_{x\in\bbZ^\nu} 
\abs{V(x)} 
\end{displaymath}
and
\begin{displaymath}
\sum_{n=1,\dots, N_\pm} [(E_n^+)^2 -4]^{1/2} + [(E_n^-)^2 -4]^{1/2} \leq 
2^{\nu-1}L_{1,\nu-1}^{c\ell} \sum_{x\in\bbZ^\nu} 
\abs{V(x)}^{1+(\nu-1)/2} .
\end{displaymath}
Simply use the induction in the dimension idea to strip off the first coordinate $x_1$ and 
then use either Theorem \ref{T5.2} or \ref{T5.3} in $\nu-1$ dimension. Of course, the 
above extension to nonconstant diagonal terms also applies.

\bigskip

\appendix
\section{The Bargmann Bound} \lb{App}
\renewcommand{\theequation}{A.\arabic{equation}}
\renewcommand{\thetheorem}{A.\arabic{theorem}}
\setcounter{theorem}{0}
\setcounter{equation}{0}

Our goal in this appendix is to prove

\begin{theorem}\lb{TA.1} Let $N(\{a\}, \{b\})$ be the number of eigenvalues of 
$J(\{a\}, \{b\})$ outside $[-2,2]$. Then
\begin{equation} \label{A.1}
N(\{a\}, \{b\}) \leq \sum_{n=1}^\infty n\abs{b_n} + (4n+2)(a_n -1)_+ 
\end{equation}
where $(x)_+ =\max(x,0)$.
\end{theorem}

This is related to a result of Geronimo \cite{Ger1, Ger2}. We provide a proof here because 
it is easy from our machinery earlier. Geronimo's second proof of this result \cite{Ger2} 
uses a Birman-Schwinger kernel as this does, but has an error in the argument that allows 
$a_n <1$ (his Lemma III.1 is wrong). Earlier papers that show $N<\infty$ if \eqref{A.1} 
holds include Geronimo-Case \cite{GC} and Chihara-Nevai \cite{CN}.

\smallskip
{\it Notes.} 1. If you translate Geronimo's result in \cite{Ger1} into our normalization 
(he has $J_0$ with $a\equiv \f12$, not $a=1$), then where we have $(4n+2)(a_n-1)_+$, he has 
$(4n+4)(a_n-1)_+ (a_n+1)$, which is weaker in two regards: $4n+2 < 4n+4$ and we have no 
$a_n+1$. We note that by looking at $b_n=0$ and $a_n=1$ for $n\geq 2$, one finds examples 
with $N=2$ and $(a_1 -1)_+$ arbitrarily close to $\sqrt{2}-1$ so that constant in front of 
$(a_1 -1)_+$ must be at least $2(\sqrt2 +1)$ and, in particular, $4n$ does not work.

\smallskip
2. We actually have separate inequalities for $N_+$ and $N_-$.

\medskip
\noindent{\bf Step 1.} $a_n\equiv 1$; $b_n \geq 0$. The proof of Bargmann's bound \cite{Barg} 
given by Birman \cite{Bir} and Schwinger \cite{Sch} works in this case. By \eqref{2.9} and the 
monotonicity with $A=J_0$, $B=J-J_0$, for $\beta >2$,
\begin{align}
\#\text{ of eigenvalues of }& A+B \geq \beta \notag \\
 &= \#\text{ of $\beta' \geq\beta$ so that $K_{\beta'}$ has eigenvalue $=1$} \notag \\
&= \#\text{ of eigenvalues of $K_\beta \geq 1$} \lb{A.2} \\
&\leq \Tra (K_\beta) \lb{A.3} \\
&\leq \Tra (K_2) \lb{A.4} 
\end{align}
where \eqref{A.2} follows from the fact that $\|K_\beta\|\downarrow 0$ as $\beta\to\infty$ 
and the strict monotonicity of the eigenvalues of $K_\beta$ noted in the proof of 
Proposition~\ref{P2.5}. \eqref{A.3} holds since
\[
\Tra(K_\beta) = \sum_{E_j^+(K_\beta)} E_j^+ \geq \sum_{E_j^+ (K_\beta)\geq 1} 
E_j^+ \geq (\#\text{ of eigenvalues of $K_\beta \geq 1$})
\]
since $K_\beta >0$. \eqref{A.4} holds since $K_\beta \leq K_2$.

The same argument that led to Proposition~\ref{P2.6} shows that $[(\beta-J_0)^{-1}]_{nm} 
=w(\beta)^{-1} \varphi_-^{(\beta)} (\min(n,m))\varphi_+^{(\beta)}(\max(n,m))$ where 
$\varphi_\pm$ solve $J_0\varphi=\beta\psi$ with $\varphi_+\in L^2$ at infinity, $\varphi_- 
(0)=0$, and $w$ is their Wronskian. As $\beta\downarrow 2$, $\varphi_+(n)\to 1$, 
$\varphi_-(n)\to n$, and their Wronskian is $1$ so
\[
(K_2)_{nm}=\min (n,m) b_n^{1/2} b_m^{1/2}
\]
and
\[
\Tra(K_2) = \sum_{n=1}^\infty nb_n,
\]
proving \eqref{A.1} in this case.

\medskip
\noindent{\bf Step 2.} $a_n \leq 1$; $b_n\geq 0$. Let $J_0(\{a_n\})$ be $J$ with $b_n=0$. 
We claim if $a_n \leq 1$ and $\beta >2$, then
\begin{equation} \label{A.5}
(\beta -J_0 (\{a_n\}))_{nm}^{-1} \leq (\beta -J_0)_{nm}^{-1}.
\end{equation}
This is a simple maximal principle argument. One first notes that if $f_n(m)=(\beta -J_0 
(\{a_n\}))_{nm}^{-1}$, then $f_n(m)>0$ (expand $(1-\beta^{-1}J)$ in a geometric series). 
Next, one notes that
\begin{align*}
((\beta - J_0)f_n)(m) &= \delta_{nm} - (1-a_{m-1}) f_n (m-1) - (1-a_m) f_n(m) \\
&\leq \delta_{nm}.
\end{align*}
Since $(\beta -J_0)^{-1}$ also has a positive matrix, applying it preserves pointwise 
matrix inequalities, so
\[
f_n(m) \leq [(\beta -J_0)^{-1} \delta_n]_m = (\beta -J_0)_{nm}^{-1},
\]
proving \eqref{A.5}.

Now \eqref{A.5} shows the Birman-Schwinger kernel for $J_0\{a_n\}$ and $J(\{a_n,b_n\})$ 
is dominated (in the sense of inequalities on matrix elements) by this for $J_0$ and 
$J(\{a_n \equiv 1, b_n\})$, so Step 1 implies
\[
\Tra (K_2 (\{a_n, b_n\})) \leq \Tra (K_2 (\{a_n \equiv 1, b_n\})) = 
\sum_{j=1}^\infty nb_n.
\]

Notice we do not have an {\it operator} inequality of the form $(\beta - J_0 
(\{a_n\}))^{-1} \leq (\beta - J_0)^{-1}$, so individual eigenvalues may not have an 
inequality (this is Geronimo's error in \cite{Ger2}).

\medskip
\noindent{\bf Step 3.} Adding $b$'s of both signs. Fix $a_n$ with $0<a_n \leq 1$. Let 
$J(\{b_n\})$ be the Jacobi matrix with $b_n$ along the diagonal and $N_\pm (\{b_n\})$ 
the number of eigenvalues $E$ with $\pm E>2$. Since $J(\{-(b_n)_-\})\leq J(\{b_n\}) \leq 
J(\{(b_n)_+\})$, we have
\[
N_\pm (\{b_n\}) \leq N_\pm (\{\pm(b_n)_\pm\}),
\]
so by \eqref{A.1} for $b_n \geq 0$ and \eqref{2.22}, we have \eqref{A.1} for the case 
$0\leq a_n \leq 1$.

\medskip
\noindent{\bf Step 4.} (General Case) Now use the idea at the start of Section~\ref{S3} 
but only for $a_n$'s with $a_n >1$. Then \eqref{A.1} holds in general since this idea reduces 
to the case $a_n\leq 1$. We use here that
\[
2n(a_n-1)_+ + 2(n+1)(a_n-1)_+ = (4n+2)(a_n-1)_+.
\]

\bigskip

\end{document}